# Wideband Spectrum Sensing at Sub-Nyquist Rates

Moshe Mishali and Yonina C. Eldar

Spectrum sensing refers to the task of identifying the frequency support of a given input signal. Standard radio-frequency (RF) lab equipment can provide this functionality. An example is a spectrum analyzer (*e.g.*, HP-8563E) which sweeps the center frequency of an analog bandpass filter and draws the in-band signal energy. The frequency support then consists of those spectrum intervals in which the signal power exceeds the noise floor. Recently, there has been growing interest in spectrum sensing for mobile cognitive radio (CR) receivers [1], which aim at utilizing unused frequency regions on an opportunistic basis. Commercialization of CR technology necessitates a spectrum sensing mechanism that reacts in real time to cognitive decisions. A mobile device, however, cannot embed solutions based on standard lab equipment, due to size, weight, power and cost limitations. Sensing in CR mobiles must be performed using minimal hardware and software resources. Therefore, enabling widespread use of CRs calls for innovative spectrum sensing techniques.

In this paper, we present a mixed analog-digital spectrum sensing method that is especially suited to the typical wideband setting of CRs. The next section briefly summarizes existing approaches to CR sensing. The advantages of our system with respect to current architectures are threefold. First, our analog front-end is fixed and does not involve scanning hardware. Second, both the analog-to-digital conversion (ADC) and the digital signal processing (DSP) rates are substantially below Nyquist. Finally, the sensing resources are shared with the reception path of the CR, so that the lowrate streaming samples can be used for communication purposes of the device, besides the sensing functionality they provide. Combining these advantages leads to a real time map of the spectrum with minimal use of mobile resources. Our approach is based on the modulated wideband converter (MWC) system [2], which samples sparse wideband inputs at sub-Nyquist rates. We report on results of hardware experiments, conducted on an MWC prototype circuit [3], which affirm fast and accurate spectrum sensing in parallel to CR communication. This can help alleviate one of the current main bottlenecks in wide-spreading deployment of CRs.

## —— *Cognitive Radios and Spectrum Sensing* ——

Traditional communication, such as television, radio stations, mobile carriers and air traffic control is carried over predetermined frequency bands. Over the years, government agencies allocated the majority of the spectrum to legacy users, reserving a particular

| | Approach | Analog front-end | ADC/DSP rate | Shared with CR reception |
|---|---|---|---|---|
| **Parametric** | Analog pilot detection / matched-filtering | scanning | n/a | ✗ |
| | Digital pilot detection / matched-filtering | ✗ | Nyquist | ✓ |
| | Cyclostationary feature extraction | ✗ | Nyquist | ✓ |
| | Waveform-based sensing | ✗ | Nyquist | ✓ |
| | Radio identification | ✗ | Nyquist | ✓ |
| **Generic** | Analog energy detection | scanning | n/a | ✗ |
| | Digital energy detection | ✗ | Nyquist | ✓ |
| | Multi taper spectrum estimation | ✗ | Nyquist | ✓ |
| | Filter bank spectrum sensing | ✗ | Nyquist | ✓ |
| | This paper | fixed | sub-Nyquist | ✓ |

[Table 1] SPECTRUM SENSING APPROACHES FOR COGNITIVE RADIO.

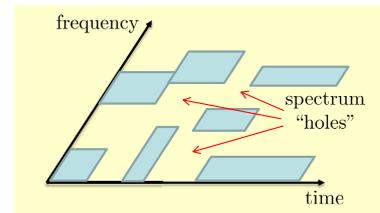

[FIG1] A cognitive radio aims at sensing the available frequency holes in consecutive time intervals.

frequency interval for each owner. This resource allocation strategy has led to spectrum congestion, to such a point that, today, the increasing demand for transmission bands can rarely be satisfied by a permanent allocation. Fortunately, studies conducted by the Federal Communications Commission (FCC) in the United States and by similar agencies in other countries indicate that the spectrum is underutilized; In a given geographical location and time duration, only a small number of legacy users transmit concurrently. This low frequency utilization, illustrated in Fig. 1, is what drives CR technology.

The idea behind CR is to exploit temporarily available spectrum holes belonging to inactive primary users. Spectrum sensing therefore takes place whenever the CR searches for available transmission holes. After a certain frequency band is chosen, the CR continuously monitors the spectrum in order to detect any change in the activity of the primary users. Once a primary user becomes active, the CR must choose another working band, or tailor its transmission to reduce in-band power. Quick and efficient spectrum sensing is evidently an essential component of CR functionality.

A special issue of the *IEEE Signal Processing Magazine* from November 2008 reviews existing CR technology [4, 5]. Current approaches for spectrum sensing are briefly summarized in Table 1 according to [6, 7]. From a bird's eye view, previous methods can be categorized into either fully hardware or fully software solutions. Known analog methods imitate the scanning mechanism used in lab equipments, thereby requiring tunable circuits, independent of the CR reception hardware. The software solutions assume that the input is sampled at the Nyquist rate

$$f_{\text{NYQ}} = 2f_{\max}, \qquad (1)$$

that is twice the highest wideband frequency $f_{\max}$. No analog preprocessing is needed and the samples can be shared with the subsequent CR stages. However, since CR typically operates in a wideband environment, the sampling rate $f_{\text{NYQ}}$ can be prohibitively large. Consequently, utilizing these sensing algorithms requires premium ADC and DSP devices that can accommodate high-rate streaming data.

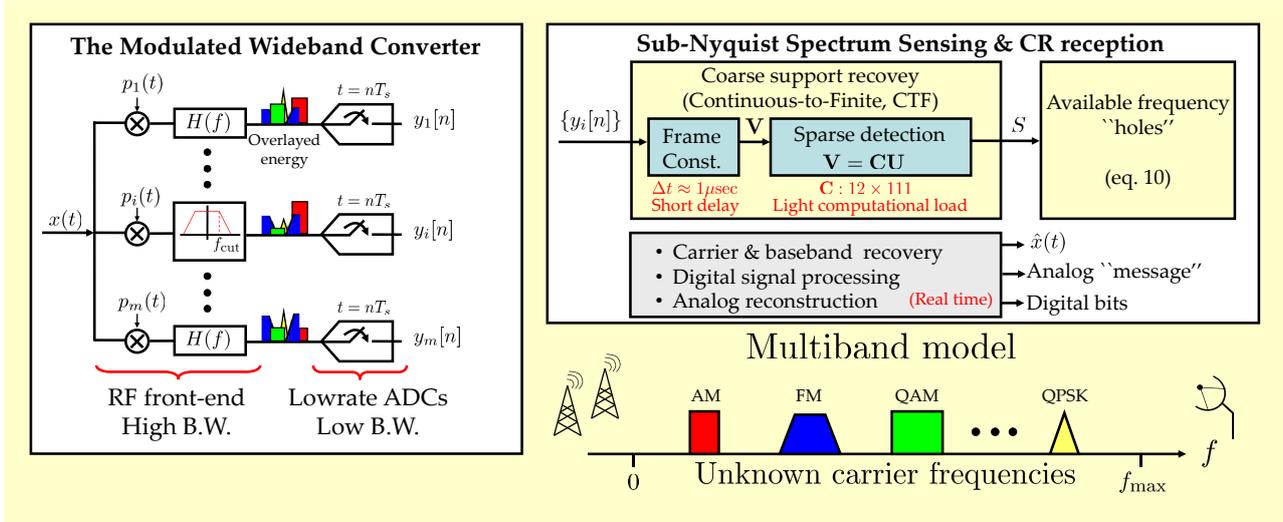

[FIG2] Block diagram of the modulated wideband converter. The digital recovery algorithm recovers the multiband input.

Table 1 distinguishes between parametric and generic approaches. Parametric methods rely on a specific structure that the input signal is assumed to obey. For example, matched filtering requires the exact transmission shape of the primary user. Other parametric approaches incorporate knowledge on preambles, midambles, synchronization bits, cyclostationarity, modulation format etc. In contrast, the generic methods avoid assumptions on the underlying signal content. Sensing based on the MWC, introduced below, belongs to the generic family of methods, and possesses additional unique features: fixed hardware, sub-Nyquist ADC and DSP rates, and shared acquisition resources between sensing and CR reception. Mixed analog-digital system design is the enabling factor behind these unique benefits.

## ——— *Sub-Nyquist Sampling: Modulated Wideband Converter* ———

Consider a signal consisting of several concurrent transmissions. In order to avoid sampling at the high Nyquist rate, the common practice in engineering is demodulation. The signal is multiplied by the carrier frequency of a band of interest, so as to shift the desired contents to the origin, where filtering and sampling at a low rate take place. When the band positions are unknown, *e.g.,* in a CR receiver, standard demodulation cannot be used.

The MWC treats multiband signals when knowledge of the carrier frequencies is present or absent. The only assumption is that the spectrum is concentrated on $N$ frequency intervals with individual widths not exceeding $B$. The sampling rate is proportional to the effective spectrum occupation $NB$ rather than $f_{\text{NYQ}}$. Typically, the spectrum is underutilized so that $NB \ll f_{\text{NYQ}}$. A digital algorithm detects the spectral support and enables either signal reconstruction or lowrate processing of the individual band contents. In this article, we take advantage of the MWC for a slightly different task – instead of aiming at the information bands, our goal is to detect the inactive support. This complementary viewpoint allows optimizing the MWC design for holes detection at the expense of the tasks that are not required in the CR settings, namely reconstruction and processing of the primary transmissions. The resulting MWC-based spectrum sensing is categorized under the "Generic" rubric of Table 1, since no assumption is made on the signal shape of the legacy users or their specific modulation techniques. Nonetheless, the sampling rate is comparable with that of a demodulator who knows the exact carrier of each transmission.

We now explain the MWC sampling stage, as depicted in Fig. 2. The system consists of a front-end of $m$ channels. In the $i$th channel, the input signal $x(t)$ is multiplied by a periodic waveform $p_i(t)$ with period $T_p$, lowpass filtered by $h(t)$, and then sampled at rate $f_s = 1/T_s$. The basic MWC configuration has

$$f_p = 1/T_p \geq B, \quad T_p = T_s, \quad m \geq 4N. \tag{2}$$

The parameter choice (2) results in

$$\text{Sampling rate} = mf_s \approx 4NB, \tag{3}$$

which in general is far below $f_{\text{NYQ}}$. In practice, an advanced configuration which we describe in the sequel is used in our hardware experiments, allowing to reduce the number of branches $m$ at the expense of increasing the sampling rate $f_s$ on each channel so that overall $mf_s \approx 4NB$.

To derive an expression for the $i$th sequence of samples $y_i[n]$ we note that since each $p_i(t)$ is periodic, it has a Fourier expansion

$$p_i(t) = \sum_{\ell=-\infty}^{\infty} c_{i\ell} e^{j2\pi f_p \ell t}, \tag{4}$$

for some coefficients $c_{i\ell}$. Denote by $z_\ell[n]$ the sequence that would have been obtained if the signal was mixed by a pure sinusoid $e^{j2\pi f_p \ell t}$ and lowpass filtered. This sequence corresponds to uniform samples at rate $f_p$ of a section of $x(t)$, conceptually obtained by bandpass filtering an $f_p$-width interval around $\ell f_p$ and demodulating to the origin. Since the system is linear, modulating by $p_i(t)$ and lowpass filtering is equivalent to summing the weighted combinations of all the sequences $z_\ell[n]$:

$$y_i[n] = \sum_{\ell=-L}^{L} c_{i\ell} z_\ell[n], \tag{5}$$

where the sum limits $-L \leq \ell \leq L$ represent the range of coefficients $c_{i\ell}$ with non-negligible amplitudes. It follows that the number of spectrum intervals that are aliased to the origin is $M = 2L + 1$. In principle, any periodic function with high-speed transitions within the period $T_p$ can be used to obtain this aliasing. One possible choice for $p_i(t)$ is a sign-alternating function, with $M$ sign intervals within the period $T_p$ [2]. Popular binary patterns, *e.g.*, the Gold or Kasami sequences, are especially suitable for the MWC [8].

Mixing by periodic waveforms aliases the spectrum to baseband, such that each frequency interval of width $f_p = 1/T_p$ receives a different weight. The energy of the various spectral intervals is overlayed at baseband, as visualized in Fig. 2. At first sight, the sequences $y_i[n]$ seem corrupted due to the deliberate aliasing. Nonetheless, the fact that only a small portion of the wideband spectrum is occupied, together with the different weights in the different channels, permits the recovery of $x(t)$. The next section explains the digital computations of Fig. 2, and in particular how the spectrum sensing functionality is achieved.

*—— Computationally-Light Software Algorithm ——*

Mathematically, the analog mixture boils down to the linear system [2]

$$\mathbf{y}[n] = \mathbf{C}\mathbf{z}[n], \tag{6}$$

where the vector $\mathbf{y}[n] = [y_1[n], \cdots, y_m[n]]^T$ collects the measurements at $t = nT_s$. The matrix $\mathbf{C}$ consists of the coefficients $c_{i\ell}$ and $\mathbf{z}[n]$ consists of the values of $z_\ell[n]$ arranged in vector form. From (2) and the definition of $z_\ell[n]$, it follows that at most $2N$ sequences $z_\ell[n]$ are active, namely contain signal energy [2]. The spectrum sensing functionality is, therefore, tantamount to finding the index set

$$S = \{\ell \,|\, z_\ell[n] \neq 0\} \tag{7}$$

which reveals the spectrum support of $x(t)$ at a resolution of $f_p$ Hz. The choice $f_p \geq B$ in (2) implies a minimal resolution which should match the expected bandwidth of legacy transmissions. Achieving smaller resolution $f_p < B$ is discussed in the next section.

Detecting $S$ by inverting $\mathbf{C}$ in (6) is not possible, since the $m \times M$ matrix $\mathbf{C}$ is underdetermined; the MWC uses $m \ll M$ to reduce the sampling rate below Nyquist. Underdetermined systems have in general infinitely many solutions. Nonetheless, under the parameter choice (2), and additional mild conditions on the waveforms $p_i(t)$, a sparse $\mathbf{z}[n]$ with at most $2N$ nonzero entries is unique and can be recovered in polynomial time [2] by relying on results in the field of compressed sensing. Further simplification of the DSP can be obtained by noting that $\mathbf{z}[n]$ are jointly sparse over time, namely the index set $S$ does not depend on the time index $n$. Therefore, $S$ can be estimated from several consecutive samples, which increases the robustness of the estimate.

Support recovery is performed in the continuous-to-finite (CTF) block in Fig. 2, which is the heart of the MWC reconstruction algorithm. The CTF builds a frame (or a basis) from the measurements using

$$\mathbf{y}[n] \xrightarrow{\text{Frame construct}} \mathbf{Q} = \sum_n \mathbf{y}[n]\mathbf{y}^H[n] \xrightarrow{\text{Decompose}} \mathbf{Q} = \mathbf{V}\mathbf{V}^H, \qquad (8)$$

where the (optional) decomposition allows removal of the noise space. The active spectrum slices are detected from the sparse solution of the following underdetermined system

$$\mathbf{V} = \mathbf{C}\mathbf{U}. \qquad (9)$$

It is proven in [9] that (9) has a unique solution matrix $\mathbf{U}$ with minimal number of non-identically zero rows, and that the locations of these rows coincide with the support set $S$ of $x(t)$. This is the point where the CR device can decide how to allocate its energy, since

$$\text{Spectrum holes} = \bigcup_{\ell \notin S} \left[ lf_p - \frac{f_p}{2}, lf_p + \frac{f_p}{2} \right]. \qquad (10)$$

Additional steps on the same sample sequences $y_i[n]$ enable processing and reconstruction of any input transmission. We refer to [2] for a detailed description of these recovery steps. Note that among the transmissions in $x(t)$, some belong to primary users while others can be CR communications. The fact that the same samples enable reconstruction of CR communications is highly important – it enables the CR to both sense the spectrum with the system of Fig. 2 and intercept communications as a standard receiver. Although beyond the current scope, we note that the CTF has a major role in signal reconstruction, beyond the robustness in estimating $S$. The CTF isolates the support recovery to a single execution of a polynomial-time algorithm. Once $S$ is known, real time processing and reconstruction is possible, that is at the (low) speed of the streaming measurements $\mathbf{y}[n]$ [10].

In the next section, we describe the circuit prototype of the MWC [3], which is used in our experiments.

## ——— *Efficient Hardware Realization* ———

The basic configuration (2) has $m \geq 4N$ channels, which may be too large to fit into a CR device. In addition, the waveforms $p_i(t)$ need to be different so as to capture linearly independent mixtures of the spectrum, which results in additional hardware per channel. To moderate the physical size, we constructed an advanced MWC configuration, proposed in [2]. In this MWC version

- the number of channels $m$ is collapsed by a factor $q > 1$ at the expense of increasing the sampling rate of each channel by the same factor, and

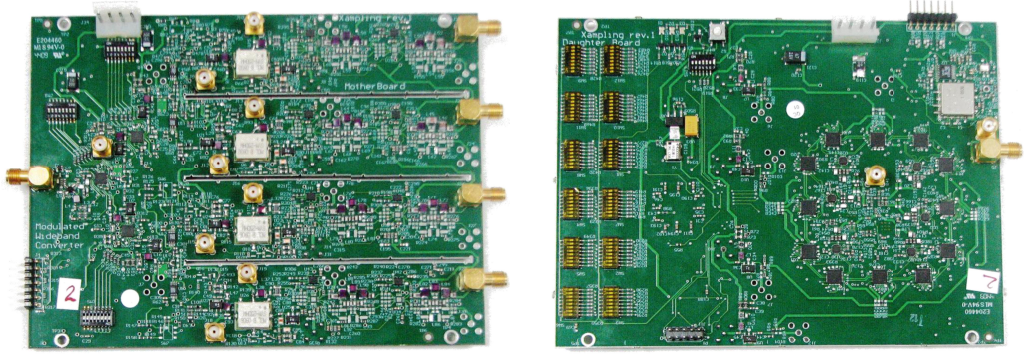

**[FIG3]** A hardware realization of the MWC consisting of two circuit boards. The left pane implements $m = 4$ sampling channels, whereas the right pane provides four sign-alternating periodic waveforms of length $M = 108$, derived from different taps of a single shift-register.

- a single shift-register provides a basic periodic pattern, from which $m$ periodic waveforms are derived using delays, that is by tapping $m$ different locations of the register.

Technically, this configuration allows to collapse channels all the way down to a single sampling channel at the same sub-Nyquist rate.

A board-level design of the MWC using this advanced configuration to treat a multiband model with $N = 6$ bands and individual widths up to $B = 20$ MHz was reported in [3]. The RF stage covers a wideband range of inputs with $f_{\text{NYQ}} = 2$ GHz and spectrum occupation $NB = 120$ MHz. An aliasing resolution of $f_p = 20$ MHz in conjunction with a sampling rate of $1/T_s = 70$ MHz results in a collapsing factor of $q = 3$. Using $m = 4$ channels, the total sampling rate is 280 MHz, which is about 14% of the Nyquist rate. Photos of the hardware are presented in Fig. 3. The resolution $f_p$ can be improved by setting $f_p < B = 20$ MHz. In the original MWC scheme [2], this choice is avoided since it increases the computations needed for signal reconstruction when a transmission occupies more than two sequences $z_\ell[n]$. The CR settings permit $f_p < B$, as only the support set $S$ is needed for sensing; Reconstructing the primary transmissions is not of interest. Since Fig. 2 is also used for CR reception, the resolution $f_p$ needs only to exceed the bandwidth of the CR communication, rather than the expected bandwidth $B$ of the primary users, which can in general be higher.

The nonordinary RF design that stems from sub-Nyquist sampling is described in [3]. For instance, lowcost analog mixers are specified for a pure sinusoid in the oscillator port, whereas the MWC requires simultaneous mixing with the many sinusoids comprising the waveforms $p_i(t)$. Another circuit challenge pertains to generating $p_i(t)$ with 2 GHz alternation rates. The severe timing constraints involved in this logic are overcome in [3] by operating commercial devices beyond their datasheet specifications. The reader is referred to [3] for further technical details.

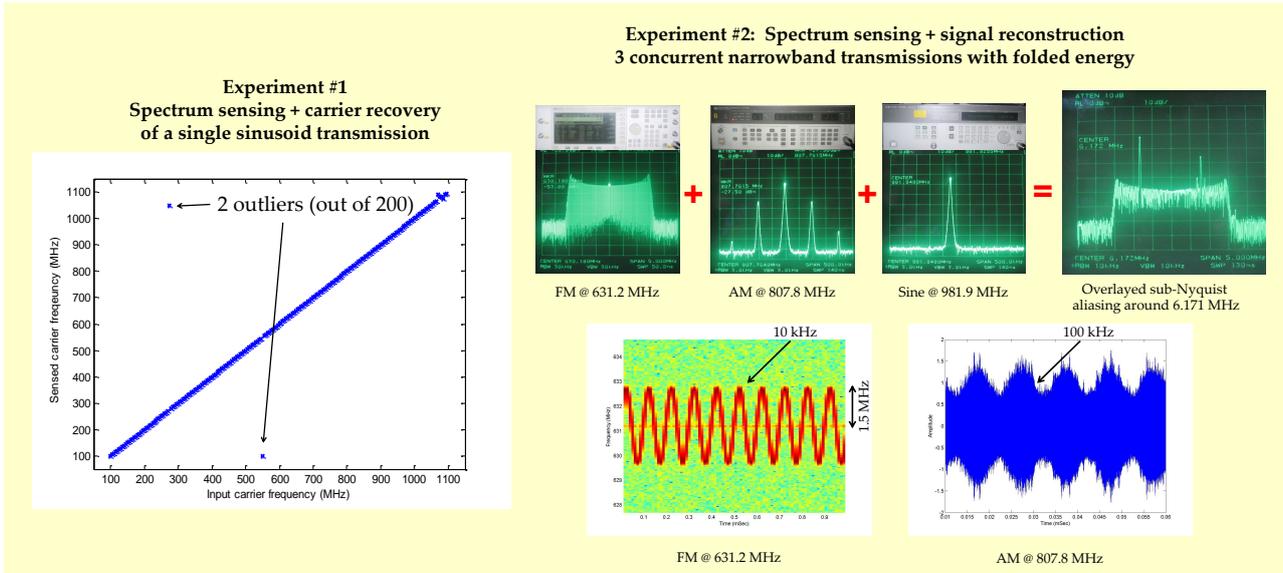

**[FIG4]** Results of hardware experiments demonstrating accurate spectrum sensing combined with signal reception, both accomplished at a sub-Nyquist rate.

## ——— *Spectrum Sensing Demonstration* ———

To verify the sensing potential of the MWC in a wideband environment, we conducted two experiments. In the first experiment, an HP-E4432B signal generator inputs a sinusoid to the MWC hardware. The four output channels were recorded using an Agilent Infiniium 54855A four-channel scope. All digital computations were carried out in Matlab. The pure sinusoid represents a challenging scenario of a legacy user with extremely narrow bandwidth. We varied the sinusoid frequency from 100 MHz to 1100 MHz in steps of 5 MHz. The CTF outputs the spectral support at resolution $f_p = 20$ MHz. We also executed the additional recovery blocks of Fig. 2 so that the algorithm estimates the input carrier frequency as well. The results, in the left pane of Fig. 4, demonstrate that out of these 200 experiments, there are only 2 outliers, which means 99% correct support and carrier estimation.

It is important to understand the reason for the outliers in Fig. 4. It is well known that finding the sparse solution of an underdetermined system, such as (9), is NP-hard. In practice, we solve (9) using polynomial-time algorithms which coincide with the true solution over a wide range of possible inputs, 99% of the cases in our experiments. The detection performance could have been improved for a higher number of sampling channels, say $m = 5$. Our design choice of a 4-channel prototype [3] represents a customary engineering compromise; saving the extra 25% in hardware size and digital computations of the $m = 5$ system, at the expense of not improving the last 1% of system performance. At this point, higher application layers can assist. For example, collaborative spectrum sensing in a network of CR devices is known to improve the overall holes detection, cf. [7].

Our lab experiments indicate an average of 10 millisecond duration for the digital computations, including the CTF support detection and the carrier estimation, measured in a standard Matlab environment. Algorithms for sparse solution of underdetermined systems consume time and typically scale with the size of **C**. The small dimensions of **C**, $12 \times 111$ in our prototype, is what makes the MWC sensing practically feasible from the computational perspective. We point out that this sensing duration is negligible with respect to cognitive protocols. For instance, the IEEE 802.22 standard for CR devices and networks, which is still under development, specifies a sensing duration of 30 seconds [11].

Interestingly, Cordeiro et al. summarized the IEEE 802.22 standard for CR in [12] in 2006 and envisioned that the sensing procedure would probably be carried out in two steps. First, a coarse and fast support detection, as does the CTF with a spectral resolution of $f_p = 20$ MHz. Then, a finer estimation, if needed. In the experiments of the previous section, the carrier recovery algorithm of [10] obtains carrier estimates within 10 kHz of the true input frequencies.

In the second experiment we exemplify the resource sharing of sensing and reception. Fig. 4 depicts the setup of three signal generators that were combined at the input terminal of the MWC prototype: an amplitude-modulated (AM) signal at 807.8 MHz with 100 kHz envelope, a frequency-modulation (FM) source at 631.2 MHz with 1.5 MHz deviation at 10 kHz rate and a pure sine waveform at 981.9 MHz. Together, this scenario represents a mixture of primary and cognitive transmitters. The carrier positions were chosen so that their aliases overlay at baseband, as the photos in Fig. 4 demonstrate. The digital recovery algorithm was executed and detected the correct support set $S$ (CTF) and input carrier positions. In addition, the figure demonstrates correct reconstruction of the AM and FM signal contents, affirming the potential of standard signal reception combined with spectrum sensing.

A video recording of these experiments and additional documentation are available online on the authors' websites:

`http://www.technion.ac.il/~moshiko/hardware.html`,

`http://webee.technion.ac.il/Sites/People/YoninaEldar/Info/hardware.html`.

In addition, we prepared a graphical package to demonstrate the MWC numerically, which is also available on our websites. The software guides the user through a four-stage flow: defining the multiband signal model $N$, $B$, $f_{\max}$, adjusting the MWC parameters and sub-Nyquist sampling, CTF support recovery, and signal reconstruction. A screenshot is showed in Fig. 5.

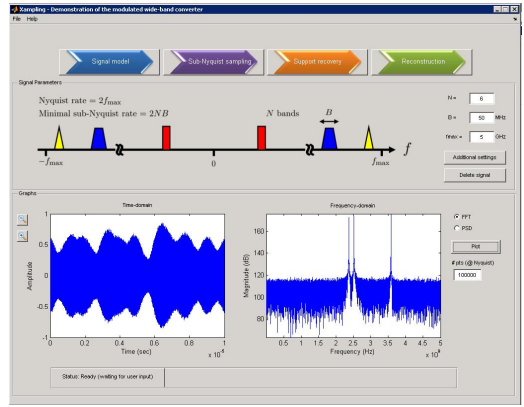

**[FIG5]** Graphical user interface of the MWC system.

## ——— *Outlook* ———

The proliferation of wireless devices necessitates flexible and efficient use of the spectrum. To render CR a widespread reality, the spectrum sensing bottleneck must be resolved. The sensing task is a crucial step in the CR life-cycle – it precedes all other cognitive decisions. Besides identifying available frequency holes, continuous monitoring is needed in order to detect appearance of primary users, an event which has immediate implications on the CR transmissions.

Cognitive communication is still a dream to come true. Research on CR is rapidly developing, providing sophisticated solutions for the multitude of challenges this technology triggers. Of the various goals, spectrum sensing is unique. In contrast to other cognitive decisions, *e.g.,* spectrum collaboration and CR networking, which are performed at higher application layers, sensing is the only task that involves the analog hardware as it begins in the analog domain. Traditionally, scanning a wide span of the spectrum is done using lab equipment, which is not constrained by size, power, cost, volume, etc. The CR era calls for innovative solutions for miniaturizing the sensing core into a mobile device, that has many other functionalities to perform in parallel. We therefore foresee much of the future research and development devoted to improving and innovating in the sensing stage.

We have demonstrated a low-rate, efficient spectrum sensing mechanism which can reliably determine inactive bands over a wide span of the spectrum, within a few milliseconds. With respect to existing sensing strategies, our approach proposes a mixed analog-digital design, with a simple and fixed analog front-end. Besides sensing, the system also serves as the CR reception path. Our design is efficiently realized in hardware and introduces only light computational loads. Hardware experiments report fast and accurate spectrum sensing due to the low sampling rate. The contribution of this work is in outlining the practical considerations for CR sensing, and providing an initial circuit-level proof of feasibility. Future work should address various hardware-related aspects, including how to miniature the design into a chip, so that it can later be embedded into existing mobile platforms.

**AUTHORS**

**Moshe Mishali** (moshiko@tx.technion.ac.il) is a PhD student of electrical engineering at the Technion, Institute of Technology, Haifa, Israel.

**Yonina Eldar** (yonina@ee.technion.ac.il) is a professor of electrical engineering at the Technion, Institute of Technology, Haifa, Israel. She is also a Research Affiliate with the Research Laboratory of Electronics at MIT, and a visiting professor at Stanford, California, United States.

## REFERENCES

[1] Mitola III, J., "Cognitive radio for flexible mobile multimedia communications," *Mobile Networks and Applications*, vol. 6, no. 5, pp. 435–441, 2001.

[2] M. Mishali and Y. C. Eldar, "From theory to practice: Sub-Nyquist sampling of sparse wideband analog signals," *IEEE J. Sel. Topics Signal Process.*, vol. 4, no. 2, pp. 375–391, Apr. 2010.

[3] M. Mishali, Y. C. Eldar, O. Dounaevsky, and E. Shoshan, "Xampling: Analog to digital at sub-Nyquist rates," *arXiv.org 0912.2495; to appear in Circuits, Devices & Systems, IET*, Dec. 2009.

[4] I. Budiarjo, H. Nikookar, and L. P. Ligthart, "Cognitive radio modulation techniques," *IEEE Signal Process. Mag.*, vol. 25, no. 6, pp. 24–34, 2008.

[5] D. Cabric, "Addressing feasibility of cognitive radios," *IEEE Signal Process. Mag.*, vol. 25, no. 6, pp. 85–93, 2008.

[6] D. D. Ariananda, M. K. Lakshmanan, and H. Nikoo, "A survey on spectrum sensing techniques for cognitive radio," in *Cognitive Radio and Advanced Spectrum Management, CogART 2009. Second International Workshop on*, MAY 2009, pp. 74–79.

[7] T. Yücek and H. Arslan, "A survey of spectrum sensing algorithms for cognitive radio applications," *Communications Surveys Tutorials, IEEE*, vol. 11, no. 1, pp. 116–130, 2009.

[8] M. Mishali and Y. C. Eldar, "Expected-RIP: Conditioning of the modulated wideband converter," in *Information Theory Workshop, 2009. ITW 2009. IEEE*, Oct. 2009, pp. 343–347.

[9] ——, "Blind multi-band signal reconstruction: Compressed sensing for analog signals," *IEEE Trans. Signal Process.*, vol. 57, no. 3, pp. 993–1009, Mar. 2009.

[10] M. Mishali, Y. C. Eldar, and A. Elron, "Xampling: Signal acquisition and processing in union of subspaces," *CCIT Report no. 747, EE Dept., Technion; arXiv.org 0911.0519*, Oct. 2009.

[11] "Standard for Wireless Regional Area Networks (WRAN) - Specific requirements - Part 22: Cognitive Wireless RAN Medium Access Control (MAC) and Physical Layer (PHY) Specifications: Policies and procedures for operation in the TV Bands, The Institute of Electrical and Electronics Engineering, Inc. Std. IEEE 802.22."

[12] C. Cordeiro, K. Challapali, D. Birru, S. Shankar *et al.*, "IEEE 802.22: An introduction to the first wireless standard based on cognitive radios," *Journal of communications*, vol. 1, no. 1, pp. 38–47, 2006.